# Temperature tunable Anderson localization for surface plasmon waves propagating in a graphene single layer placed on a random InAs grating

**Abbas Ghasempour Ardakani[1,*], Marzieh Sedaghat Nejad[1]**

*[1]Department of Physics, Shiraz University, Shiraz 71454, Iran*
*Corresponding author: aghasempour@shirazu.ac.ir*

In this paper, we propose a one-dimensional disordered plasmonic structure composed of a graphene single layer placed on a random grating composed of InAs. The propagation of a plasmonic wave through this structure is investigated numerically. By calculation of normalized localization length for systems with different disorder strengths, it is determined whether or not the system is in the localized regime. For some frequencies, depending on the disorder level, Anderson localization occurs for plasmonic waves propagating through the graphene layer. Furthermore, the effect of optical loss on the localization length is studied. By calculating the localization length at different temperatures, it is observed that Anderson localization of graphene plasmons is temperature dependent and can be controlled by changing the temperature. In the transmission spectrum for each random realization, there are some resonance peaks which are blue-shifted with increasing the temperature. Finally, the effects of Fermi energy level of the graphene layer and width of air gaps on the individual transmission resonances are examined.

## 1. INTRODUCTION

Surface plasmon polaritons (SPPs) are electromagnetic surface waves propagating at the interface between dielectrics and metals or semiconductors in which the electronic charge carriers can move freely. These surface waves result from the coupling between photons and plasma oscillations at the surface of metals or semiconductors [1]. Because SPPs allow the confinement and propagation of electromagnetic energy in the sub-wavelength scales, they have potential applications in fabrication of nonostructure devices for optical communications. So far different metal based plasmonic devices such as optical waveguides [2], filters [3], all-optical switchers [4], lenses [5], and Bragg reflectors [6], have been designed which operate in the visible and near-infrared regions.

Anderson localization is a physical phenomenon which is observed in disordered systems [7]. This effect results from the interference effect and was predicted for the first time by Anderson for spin diffusion and electronic conductance in disordered solids [8]. Anderson localization has been demonstrated for different waves such as acoustic waves [9], matter waves [10] and electromagnetic waves over different frequency ranges [11-18].

So far, few works have addressed the Anderson localization for surface plasmon polariton waves. Mcgurn et al studied the resonant scattering of a transverse magnetic (TM) polarized light on a randomly rough grating created on a metallic surface [19]. They found that the localization of surface plasmon polaritons resulted from surface roughness causes the angular dependence of the intensity of the nonspecularly reflected light to become maximum in the antispecular direction. Arya et al demonstrated that localization of surface plasmon polaritons (SPP) occurs parallel to the metal surface in the presence of random roughness only in a certain frequency range [20]. They obtained a self consistent equation by using the diagrammatic method to calculate the renormalized diffusion coefficient of SPPs. They indicated that the enhancement factor of localized SPPs is always larger than that of extended SPPs. Pincemin et al studied the propagation of SPPs along a finite grating using a volume integral equation [21]. They showed that band gaps appear in the transmission spectrum for some frequency ranges. They observed some oscillations in the transmission of the structure which was due to finite size effects. In addition, they introduced a defect in the grating and found that the defect mode appears as a narrow peak within the band gap and enhancement of near field. Finally, they investigated the transmission of SPPs in random gratings that are periodic on average. They found that the width of band gap decreases in the presence of randomness. The reflection of finite grating increases with increase of its length for frequencies out of the band gap but near the band edge. This behavior confirms that Anderson localization occurs for SPP waves. Shi et al demonstrated light localization in a random array of coupled metallic nanowires at a deep-subwavelength scale [22]. The disorder is introduced in this array of plasmonic nanowires by random variation of nanowire radius. The size of localized state in this structure is much smaller than the optical wavelength. They showed that the gain needed for compensation of losses in the plasmonic components of this system is lower than loss coefficient of metals due to the presence of Anderson localization. Maradudin et al studied both analytically and numerically the propagation of SPPs along a metal surface on which a random roughness is created with a finite length [23]. They indicated that when the roughness strength increases, the localization length becomes smaller than the vacuum wavelength of SPP waves. As a result, SPP waves become localized when the disorder level of surface roughness is larger than a certain value. Optical loss in metals and difficulty in control of their permittivity limit the performance of plasmonic devices based on metals. In recent years, the graphene has been introduced as a strong candidate for propagation of SPPs [24]. Graphene as a two dimensional honey-comb lattice of carbon atoms has attracted considerable interest due to its exceptional electronic and photonic properties [24].

In recent years, it is demonstrated that graphene can be a promising platform to support SPPs [24]. SPPs in graphene compared to metals exhibit appealing features such as tunability by doping and gating [25], high optical field confinement [26] and low ohmic losses [24]. Recently, graphene plasmon polariton (GPP) is used to design different structures such as waveguides [27], modulators [28], filters [29], photodetectors [30], and sensors [31].

In recent years, plasmonic Bragg reflectors are designed and investigated based on metal-insulator-metal structures in which periodic grooves are created symmetrically [32,33]. In ref [32] a periodic structure in form of Metal-Insulator-Metal (MIM) waveguide has been proposed and it has been shown when the two kinds of dielectric material with a high refractive index and different widths alternately put together, the photonic band gap is observed. In addition, a SPP nanocavity is formed by breaking the periodicity of the proposed structure. Tao et al proposed a plasmonic Bragg reflector structure formed in graphene waveguides and investigate its performance. They showed that the plasmonic Bragg reflector can produce a broadband fast-tunable stopband. By introducing a defect in the Bragg reflector, they obtained a defect resonance mode with a high-Q factor [34]. Luo et al suggested a narrow band plasmonic filter using graphene waveguide with asymmetrical structure composed of two Bragg reflectors with different air trench sizes [35]. This structure provides a transmission peak with bandwidth of 0.12 μm in the mid-infrared region around the wavelength of 6.9 μm. Zhuang et al proposed a structure composed of periodically arranged graphene nanoribbon waveguides with different widths that provides the needed refractive index contrast to obtain Bragg reflectors without exploiting different substrates [36].

In 2016 Shi et al designed a tunable band-stop filter based on GPs with periodically modulated chemical potentials resulting from varying of the chemical potential of graphene [37]. Li et al proposed a novel line-shaped plasmonic resonator formed by a single graphene sheet deposited on two air slits with different widths [38]. It was demonstrated such structure acted as a band-pass filter in the mid-infrared region. In addition, the transmission spectrum can be tuned with changing the length of the graphene strip sandwiched in cavities as well as a small change in the chemical potential of graphene. Gao et al designed a tunable plasmonic filter made of a graphene split ring (GSR) resonator coupled with a graphene nano-ribbon and analyzed it numerically by the finite element method (FEM). The transmission spectrum of the proposed filter can be changed by tuning the shape, orientation, and Fermi level of GSR [39]. In our previous work, we propose a temperature tunable one dimensional plasmonic photonic crystal based on a graphene single layer deposited on an InAS grating whose groves are filled with air [40]. Our numerical results in this paper demonstrated that the stop band in this structure can be easily tuned by changing the ambient temperature and it is blue shifted with increase of temperature. To achieve a temperature tunable Fabry-Perot microcavity, we introduced a defect into the proposed structure. It was found that the defect mode moves to higher frequencies with increase of temperature as well as increase of Fermi energy level. Recently, Sani et al have used a 1D random structure based on silicon and air grooves for excitation of SPPs in a graphene monolayer [41]. They demonstrated that graphene plasmon can be excited in the case of disordered structure lying under the graphene layer and they showed that the intensity of graphene plasmon enhanced in some regions along the graphene layer. The authors mentioned that this effect results from Anderson localization for plasmon waves. It is well known that to understand whether a system is in the localized regime or not, the calculation of the localization length is necessary which requires ensemble averaging. But there exist no calculations of localization length in ref [41].

In this paper, we propose a structure in which a single graphene layer is deposited on a one-dimensional disordered multilayered structure composed of air and InAs semiconductor material alternatively. Due to the presence of disorder in the 1D structure located under the graphene layer, the plasmonic wave propagating in the graphene becomes localized. Because the electric permittivity of InAs significantly depends on the temperature, we expect that Anderson localization for plasmonic wave propagating in the graphene layer is affected by the ambient temperature.

The remaining parts of the paper are organized as follows. Section 2 is devoted to the design of structure and the simulation method. We present the numerical results and discussion in section 3. Finally, the manuscript is finished with some conclusion in section 4.

## 2. THE DESIGNED STRUCTURE AND SIMULATION

In the present work, we first consider a graphene single layer deposited on a disordered one-dimensional structure composed of a InAs substrate into which air trenches are created randomly with the same widths. This disordered plasmonic structure is schematically shown in Fig. 1. In fact, in this structure, the graphene layer is placed on a 1D random system composed of binary layers made of two materials with different dielectric constants. It is assumed that the structure is embedded in air. In this figure, the green layer simulates the graphene layer, while red and blue layers simulate InAs layer and air trenches, respectively. In Fig. 1, the width of air gaps and InAs layers are shown with $d_{air}$ and $d_{InAs}$, respectively. $d_{air}$ is taken to be a fixed number, while $d_{InAs}$ are chosen to be random numbers as $d_{InAs}$ =$d_0(1+\delta)$ where $d_0$ is the mean width and $\delta$ are random numbers distributed uniformly in the range [-q, q], so q indicate the disorder strength. In Fig. 1, the thickness of InAs substrate and the depth of each air trench are shown with D and h, respectively.

To obtain the transmission spectrum for graphene plasmon propagating through the random structure, we need to compute the optical properties of graphene which are described by a surface conductivity $\sigma_g$. In the THz frequency region, $\sigma_g$ which is quantitatively described by the Kubo formula as $\sigma_g = \sigma_{intra} + \sigma_{inter}$. The first term corresponds to the intraband electron-photon scattering and is given by [42,43]:

$$\sigma_{intra} = \left( \frac{2ie^2 k_B T}{\pi \hbar^2 \left( 2\pi f + \frac{i}{\tau} \right)} \right) \ln \left( 2\cosh \left( \frac{E_f}{2k_B T} \right) \right) \qquad (1)$$

and the second term corresponds to the interband transition contribution:

$$\sigma_{inter} = \left(\frac{e^2}{4}\right)\left(\frac{1}{2} + \frac{1}{\pi}\tan^{-1}\left(\frac{2\pi f - 2E_f}{2k_BT}\right)\right)$$

$$-\left(\frac{i}{2\pi}\right)\ln\frac{(2\pi f + 2E_f)^2}{(2\pi f - 2e_f)^2 + (2k_BT)^2} \qquad (2)$$

where e is the electron charge, $k_B$ is the Boltzmann's constant, T is the ambient temperature, $= \frac{h}{2\pi}$ is the reduced Plank constant, $f$ is the frequency, $\tau = \frac{E_f\mu}{ev_f^2}$ is the relaxation time which depends on the carrier mobility $\mu$ (here is set as 1), Fermi velocity of the graphene $v_f = 10^6\frac{m}{s}$ and the Fermi energy $E_f = v_f\sqrt{\pi n_{2D}}$. The Fermi energy depends on the charge density $n_{2D}$ and can be controlled by chemical doping or applying a bias voltage. The optical properties of graphene are usually described with the equivalent permittivity of graphene as $\varepsilon_g = 2.5 + i\frac{\sigma_g}{2\pi f\varepsilon_0 d_g}$, where $\varepsilon_0$ is the vacuum permittivity and $d_g$ is the thickness of graphene monolayer [34]. In our simulation the graphene sheet is treated as an ultra thin film with a thickness of $d_g = 1nm$ and dielectric permittivity of $\varepsilon_g$ introduced above.

The corresponding dispersion relation for TM polarized SPPs propagating in a graphene sheet sandwiched between two dielectric media with dielectric permittivity of $\varepsilon_1$ and $\varepsilon_2$, can be described as [44]:

$$\frac{\varepsilon_1}{\kappa_1} + \frac{\varepsilon_2}{\kappa_2} + i\frac{\sigma_g}{\varepsilon_0\omega} = 0 \qquad (3)$$

where

$$\kappa_m^2 = q^2 - \frac{\omega^2\varepsilon_m}{c^2}, (m = 1,2) \qquad (4)$$

and

$$q = n_{eff}k_0 = n_{eff}\frac{2\pi}{\lambda} \qquad (5)$$

Here, q and $n_{eff}$ denote the constant of propagation and effective refractive index for SPPs, and $\lambda$ is the wavelength of incident light in free space while $k_0$ is its wave number. According to Eq. (3-5), the effective refractive index for SPPs depends on the media surrounding the graphene sheet. Therefore, we expect that if a graphene sheets is placed on a random system, SPPs encounter a random effective refractive index and Anderson localization effect occurs for them. In addition, when refractive index of one of surrounding dielectric materials is sensitive to some external parameters, we expect that the Anderson localization for graphene plasmon can be manipulated with these external parameters.

For the THz frequencies the permittivity of InAs is given by the simple Drude model [45]:

$$\varepsilon_s = \varepsilon_\infty - \frac{\omega_p^2}{\omega^2 + i\gamma\omega} \qquad (6)$$

where $\varepsilon_\infty$ is the high frequency permittivity, $\gamma$ is the damping constant. The plasma frequency is given by $\omega_p = \sqrt{\frac{Ne^2}{\varepsilon_0 m^*}}$ which depends on the intrinsic carrier density N, the effective mass $m^*$ of free carriers, the electric charge e and the free-space permittivity $\varepsilon_0$ .The damping constant $\gamma$ is $1.4\times10^{13}(s^{-1})$ [45,46],. The intrinsic carrier density N (in $m^{-3}$) in InAs, for temperature range 350 K<T<900 K, can be written as:

$$N = 2.14\times10^{15}T^{\frac{3}{2}}\exp(-\frac{0.47}{2k_BT}) \qquad (7)$$

where $k_B$ is the Boltzmann's constant, the temperature T is in Kelvin and $k_BT$ is in eV. The other parameters are set as $\varepsilon_\infty = 12.25$, $m^* = 0.022m_e$ and $m_e = 9\times10^{-31}kg$ [45, 46]. According to Eq. (7), N is temperature dependent, resulting the plasma frequency $\omega_p$ of InAs strongly depends on the temperature T and can be tuned by changing the temperature T. Consequently, the relative permittivity of InAs in Eq. (6), compared to metals, is very sensitive to the temperature in the mid-infrared region. Therefore, if one of the constituents of the 1D disordered structure lying the graphene layer is InAs, we expect that Anderson localization for graphene plasmons is tunable with changing the temperature.

To quantify the localization behavior in 1D random system, the localization length should be calculated. For a localized state, the localization length $\xi$ decreases exponentially with the system length L and is defined as [47]:

$$\xi = -\lim_{L\to\infty}\frac{2L}{\ln(T_R)} \qquad (8)$$

where $T_R$ is Transmittance. In the case of sufficiently long 1D random structures, the obtained $\xi$ from the above equation is well-defined due to self-averaging. However, for the cases at which the random system has a finite length, the localization length is computed by ensemble averaging of the transmittance T over many random realizations as follows:

$$\xi = -\frac{2L}{\langle\ln(T_R)\rangle} \qquad (9)$$

where <...> denotes the ensemble averaging. In the case of lossy random structures, both Anderson localization and optical absorption are responsible for the exponential decay of wave intensity through the structure. Therefore, in this paper we define the attenuation length as follows to account for absorption:

$$l_a = -\frac{2L}{\langle\ln(T_R)\rangle} \qquad (10)$$

which is smaller that localization length $\xi$. The proposed structure is investigated numerically by 2D simulations using finite-element method software (COMSOL Multiphysics). In the FEM simulations, the number of layers of 1D random structure placed under the single graphene layer is chosen to be N=40 which are alternately arranged in the y direction. In order to calculate the transmission of the structure, we use the numeric port to excite SPPs in graphene for different frequencies. In COMSOL, one can accurately find the boundary mode near the given effective refractive index by running boundary mode analysis [48].

The values of those parameters used in our simulations are taken to be $d_{air}$=100 nm, $d_{I}$=40 nm, h=25 nm, D=100 nm, and $\varepsilon_{air}$ = 1 .

## 3. Results and discussion

We consider a plasmon wave incident from the left onto the proposed random structure as indicated by the yellow arrow in Fig. 1. First, we calculate the normalized localization length ($\xi$/<L> where <L> is the ensemble average of the random system length.) of the proposed structure versus frequency in the frequency range 30 THz<v<60 THz for different disorder levels q=0, 0.1, 0.3, 0.5 and 0.7 and show the corresponding results in Fig. 2. The case q=0 corresponds to the periodic system. The ambient temperature and Fermi energy level are taken to be T=700 K and 0.7 eV, respectively. In Fig. 2, we neglect the imaginary part of dielectric permittivity of graphene and InAs and then consider the effect of absorption of these materials on the localization length. It is well known that in the lossless disordered system, for frequencies at which the normalized localization length is smaller than one, the system is in the localized regime. However, when normalized localization length is larger than one, the system is in the extended regime. It is found in Fig. 2 that when the one-dimensional plasmonic structure is periodic (q=0), there are three photonic band gaps and two transmission bands in the structure. For the transmission bands the system is in the extended regime while for stop bands the system is in the localized regime. As disorder level increases to q=0.1, the normalized localization length at both transmission and stop bands corresponding to the case q=0 decreases in most of frequencies. It is clearly seen that at some frequencies located in the transmission bands corresponding to the periodic structure the system is in the extended regime. With increasing the disorder level to q=0.3, the normalized localization length in transmission bands decreases while the normalized localization length in the stop bands increases. For this disorder level all frequencies located in the range 30 THz<v<60 THz are in the localized regime. With further increase of disorder level to q=0.7, the same effects are observed and all frequencies are in the localized regime. The case q=0.5 is not displayed in Fig. 2, because for this disorder strength we consider the effects of ambient temperature on the localization length next. As shown in Fig. 2, with increasing the disorder level in the proposed one dimensional plasmonic structure, the number of localized plasmonic states increases with increasing the disorder level as for propagation of other electromagnetic waves in random structures. Therefore, our results in this paper confirm that surface plasmon waves propagating in a graphene layer can be Anderson localized provided the graphene layer is deposited on a 1D disordered structure and the disorder level is larger than a certain value.

Now we study the effect of ambient temperature on the localization length and Anderson localization behavior. So, we calculate localization length versus frequency at different temperatures T=700, 750, 800 and 850 K and show the corresponding results in Fig. 3. We set the disorder level to be q=0.5 and the other parameters are the same as those in Fig. 2. Because the variation of real and imaginary parts of relative permittivity of InAs with temperature is stronger at higher temperatures compared to lower ones [40], here we consider only the effects of temperatures higher than 700 K. As shown in Fig. 3, the localization length significantly depends on the temperature. One can see that at most frequencies the localization length increases with increasing the ambient temperature. Furthermore, the number of extended states for temperature T=850 K is larger compared to the lower temperatures. This effect is due to the fact that with increase of temperature the refractive index contrast between effective indices of air/graphene/air and air/graphene/InAs waveguides decreases. Anderson localization becomes stronger with increase of refractive index contrast due to the enhancement of wave scattering. It should be noted that the fluctuation in the localization length decreases with increasing the number of random realizations. However, further increase of random realization is very time consuming. So our results here confirm that the localization behavior of graphene plasmon in the proposed structure can be controlled with changing the ambient temperature. The manipulation of surface plasmons in such disordered structures is of great importance from practical point of view.

To understand how the loss in the graphene monolayer and InAs material affects the localization length, we calculate the attenuation length defined in Eq. (10) for structures in Fig. 3 with taking into account the imaginary part of relative permittivity of graphene layer and InAs material in the simulations. Fig. 4 shows the corresponding results for different ambient temperatures T=700, 750, 800 and 850 K. The other parameters are the same as those in Fig. 3. One can see from this figure that in the presence of optical loss, the attenuation length is smaller than corresponding localization length in Fig. 3. This result is reasonable because both Anderson localization and absorption are responsible for exponential decaying of the surface plasmon energy in graphene.

The localization length which is obtained by ensemble averaging describes the exponential decay of the graphene plasmon wave propagating in an infinite 1D disordered structure. But in the case of finite random systems, the exponential decay is observed only for some realizations. Any realization of random systems with a finite length supports resonance wavelengths with a high transmission. In fact existence of these resonance transmissions confirms the excitation of localized modes in the disordered plasmonic structure. In recent years, it is investigated how absorption [49,50], nonlinearity [51] and magnetooptic effect [52] change the location and peak of transmission resonances for electromagnetic waves in random systems. Here, we study the effects of ambient temperature on the transmission resonances for the graphene plasmon propagating in the random structure. In recent years, Rüting et al studied how SPPs propagate in dielectric loaded waveguides in which randomly placed scatterers are introduced [53]. However in this reference the propagation of SPPs occurs in the interface between a metal and surrounding dielectrics. The transmission of SPPs through the proposed structure shows a number of peaks which arise from single and multiple resonances. Some of these extended states are called necklace states which appear when two or more resonances have the same frequency. To investigate the effects of temperature on the transmission resonances, we calculate the transmittance for the graphene plasmon wave propagating through the structure at different ambient temperatures T=700, 750, 800, 850 K for a random realization and show the corresponding results in Fig. 5. The strength of randomness in this structure is taken to be q=0.5 and the other parameters are the same as those in Fig. 2. In addition, the transmittance for the corresponding periodic structure with $d_{InAs}$=40 nm is also shown in Fig. 6. One can see that with introduction of randomness some resonance peaks appear in the transmission bands and band edges. It is clearly seen that the resonance peaks at v=34.80 and 36.15 THz are blue shifted with increasing the temperature. In Fig. 8 of our previous work [40], $n_{eff}$ of SPPs propagating at graphene single layer sandwiched between air and InAs versus frequency are displayed at different temperatures. It is found from this figure that at each frequency the increase of temperature leads to the reduction of $n_{eff}$. According to Bragg relation the center frequency of stop and transmission bands are shifted to higher frequencies with increasing temperature. Because the transmission peaks in the disordered system appears in the transmission bands and band edges corresponding to the periodic structure, we expect that the resonance peaks are blue-shifted with increasing the temperature.

These resonance peaks suggest that the proposed disordered plasmonic structure in this paper can be used as plasmonic filters. Furthermore, these structures are suitable for observation of nonlinear optical effects provided the optical nonlinear materials are located in the random localization centers.

We also display the distribution of $H_z$ field along the random structure for resonance peaks f=30.45, 36.30 and 40.35 THz at T=800K in Fig. 7. It is clearly seen that at each resonance frequency, there are some regions with high intensities in the random structure with random positions. The position of these regions significantly depends on the resonance wavelength.

Next, we investigate the effect of Fermi energy on the resonance peaks. Fig 8 shows the transmittance of graphene plasmon for different Fermi energies at a fixed temperature T=800 K. It is found from this figure that the transmission spectrum significantly depends on the Fermi energy. As shown in Fig. 8, with increasing $E_F$ from 0.5 to 0.9 eV, the resonance peaks are blue shifted. This effect is due to increase of Bragg frequency with increasing Fermi energy.

Now, we study how the widths of air gaps under the graphene layer affect the transmission of the plasmon wave through the disordered plasmonic structure shown in Fig. 1. To this end, we increase $d_{air}$ from 100 nm to 150 nm and calculate the transmittance of the structure at different temperatures T=700, 750, 800, 850 K. The other parameters are the same as those in Fig. 5. We show the corresponding results in Fig. 9. As shown in this figure, the resonance peaks are blue shifted with increase of temperature as in Fig. 5. However, the transmittance values at resonance peaks enhance in Fig. 9 compared to Fig. 5 with increasing $d_{air}$.

Consequently, in this paper we show that Anderson localization can occur for graphene plasmons using the calculation of normalized localization length. In addition, by introduction of InAs layers with random thicknesses in the graphene based 1D plasmonic structures, temperature tunable Anderson localization for surface plasmon polaritons propagating through the graphene monolayer is achieved. In the case of disordered plasmonic structure, some resonance peaks appear in the transmission spectrum which can be used for fabrication of plasmonic filters. The operating frequency of these filters can be controlled by changing the ambient temperature as well as Fermi energy level of graphene. It should be noted that in ref [41] no attempts were performed for calculation of localization length and the top of structure is illuminated by the incident wave for excitation of surface plasmons in graphene. Here, we demonstrate that the system is in the localized regime based on the value of localization length relative to the length of system.

# 3.  4. CONCLUSION

We have studied Anderson localization behavior in a 1D plasmonic disordered system made of a single graphene layer deposited on a one-dimensional disordered multilayered structure composed of air and InAs semiconductor material alternatively. By employing the 2D FEM simulation, the localization length for surface plasmon waves propagating through the graphene layer is calculated for different disorder levels. It has been found that by increasing the disorder strength the number of localized states increases. Because the relative permittivity of InAs is very sensitive to the temperature in the mid-infrared region, it has been demonstrated that the localization behavior of graphene plasmon in this structure can be controlled with changing the ambient temperature. In addition, by taking into account the optical loss in the graphene layer and InAs material, one can see that the localization length decreases because both Anderson localization and absorption are responsible for the exponential field decay. The transmittance of surface plasmon waves is calculated for a random realization at different temperatures and one can see that some resonance peaks appear in the spectrum. When temperature increases, these transmission peaks shifts to higher frequencies. The calculation of $H_z$ field distribution at resonance peaks confirms the formation of random cavities in the structure. At a fixed temperature, as the Fermi energy level of graphene increases, the transmission peaks in the mid-infrared region are blue-shifted. With increase of the width of air gaps, we have observed that the resonance peaks are blue-shifted with increases of temperature again while the values of transmission enhances compared to the case with narrower air gaps.

## Figure Captions

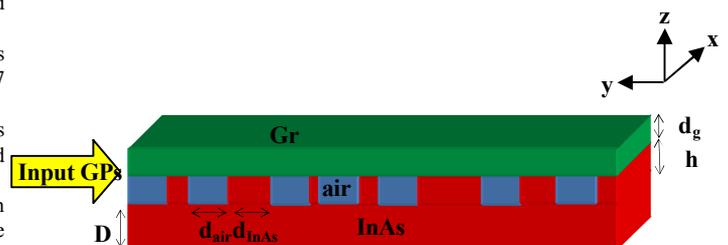

Fig. 1. Schematic of 1D disordered plasmonic structure based on a graphene sheet placed on an InAs substrate perforated with air grooves randomly. The SPP wave is excited from the left side of the structure.

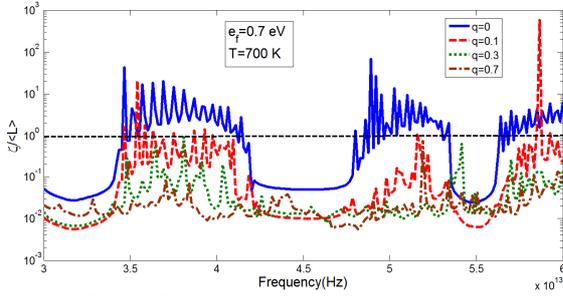

Fig. 2. Normalized localization lengths of the 1D disordered plasmonic structure versus frequency for different disorder levels q=0, 0.1, 0.7 and 0.9. The values of Fermi energy, temperature and thickness of air gaps are ef= 0.7 eV, T=700 K and d$_{gap}$= 100 nm, respectively.

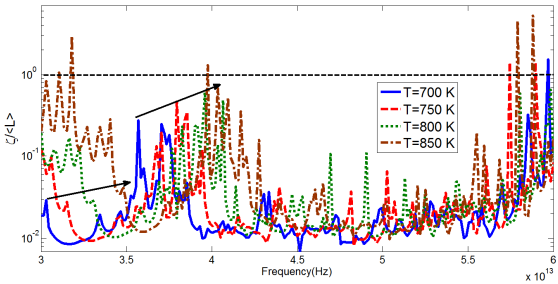

Fig. 3. Normalized localization lengths of the 1D disordered plasmonic structure versus frequency for different temperatures T=700, 750, 800 and 850 K. The other parameters are the same as those in Fig. 2.

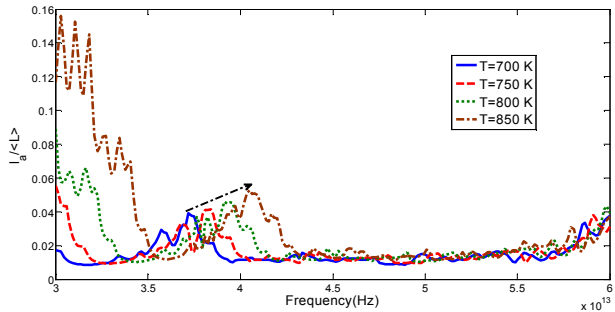

Fig. 4. Normalized attenuation lengths of the 1D disordered plasmonic structure versus frequency for different temperatures T=700, 750, 800 and 850 K. The other parameters are the same as those in Fig. 2.

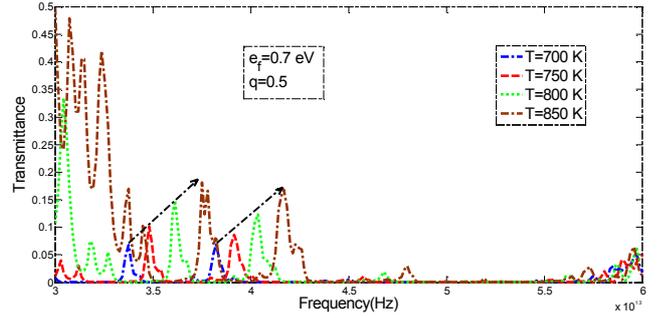

Fig. 5. Transmittance for a random realization of 1D disordered plasmonic structure with the strength of randomness q=0.5, for different temperatures T= 700, 750, 800 and 850 K. The other parameters are the same as those in Fig. 2.

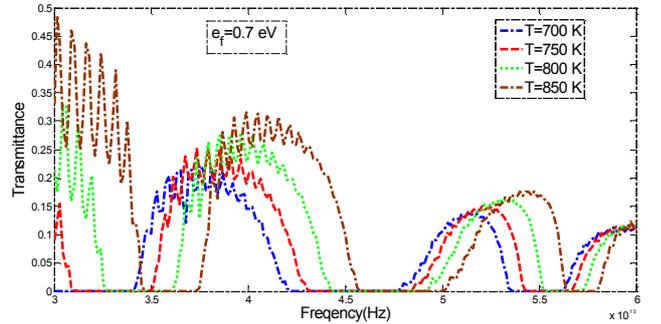

Fig. 6. Calculated transmittance for a 1D periodic plasmonic photonic crystal with d$_{gap}$= 40 nm (q=0) at different temperatures T= 700, 750, 800 and 850 K. The other parameters are the same as those in Fig. 2.

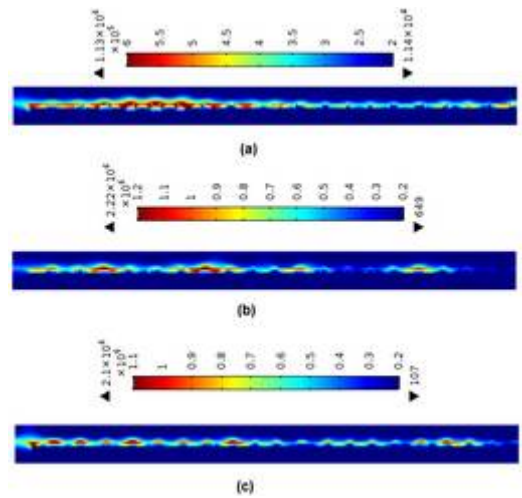

Fig. 7. H$_z$ field distribution in the 1D disordered plasmonic structure corresponding to Fig. 5 at T=800 K and e$_f$ = 0.7 eV for resonance peaks with frequency of (a) 30.45 THz, (b) 36.30 THz and (c) 40.35 THz.

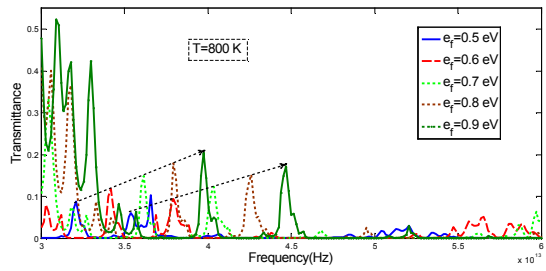

Fig. 8. Transmission spectra for a random realization of 1D disordered plasmonic structure at T= 800K for different Fermi energies $e_f$ = 0.5, 0.6, 0.7, 0.8 and 0.9 eV. The other parameters are the same as those in Fig. 2.

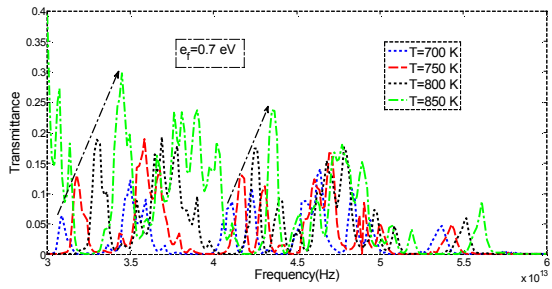